\documentclass[preprint]{ptephy_v1}

\preprintnumber{XXXX-XXXX} 

\usepackage{amsmath} 
\usepackage{graphics} 
\usepackage{bm} 

\begin{document}

\title{
Muon-Spin Motion at the Crossover Regime between Gaussian and Lorentzian Distribution of Magnetic Fields
}

\author{Muhamad Darwis Umar$^{1,2}$}
\author{Katsuhiko Ishida$^1$}
\author{Rie Murayama$^1$}
\author{Dita Puspita Sari$^{1,3}$}
\author{Utami Widyaiswari$^{1,4}$}
\author{Marco Fronzi$^3$}
\author{Harion Rozak$^{1,5}$}
\author{Wan Nurfadhilah Zaharim$^{1,5}$}
\author{Isao Watanabe$^{1,2,4,5}$}
\author{Masahiko Iwasaki$^1$\thanks{These authors contributed equally to this work}}
\affil{$^1$RIKEN, Hirosawa 2-1, Wako-shi, Saitama 351-0198, Japan, \email{Ishida@riken.jp,nabedon@riken.jp}} 
\affil{$^2$Hokkaido University, Kita 10 Nishi 8, Sapporo, 060-0810, Japan,}
\affil{$^3$Shibaura Institute of Technology, 307 Fukasaku, Minuma, Saitama 337-8570, Japan,}
\affil{$^4$Universitus Indonesia, Depok 16424, Indonesia.}  
\affil{$^5$Universiti Sains Malaysia, Pulau Pinang 11800, Malaysia}

%
%
%


\begin{abstract}%
The muon spin relaxation method ($\mu$SR) is a powerful microscopic tool to probe electronic states of materials observing local magnetic field distributions on the muon. It often happens that a distribution of local magnetic fields shows intermediate state between Gaussian and Lorentzian shapes. In order to generally describe intermediate field distributions, we dealt the convolution of two isotropic distributions in the three dimension and derived exact muon-spin relaxation functions which can be applied to all crossover regimes between the Gaussian and Lorentzian. 
\end{abstract}

\subjectindex{xxxx, xxx}

\maketitle

\section{Introduction}

\ \ \ The muon spin relaxation ($\mu$SR) technique is a powerful microscopic probe to investigate electronic states of materials on the atomic view point. After the muon is injected into target materials keeping the self-spin polarization along the injection trajectory, the muon decays to a positron with the lifetime of 2.2 $\mu$sec, while interacting with surrounding electronic states.\cite{MuonPhys} The asymmetric emission of positrons along the muon-spin direction is detected  by forward and backward counters and the time dependence of the muon-spin polarization ($\mu$SR time spectrum) is measured. By analyzing the $\mu$SR time spectrum by applying some analysis functions, we can investigate magnetic transitions,\cite{UemuraLCO,Budnick,IW1,IW2,IW3,Zheng,O2,Pattenden,Aidy} supeconducting states,\cite{Luke,AidyTRSB,UemuraPlot,Alan,AdachiZnNi} molecular motions,\cite{Blundell} electronic orbital states,\cite{Ian,Fadhilah} ionic/spin diffusions\cite{Nagamine,Ishida,IW4,Sugiyama,Martin,PrattDif} and so on. Since the muon is injected from the outside to materials, $\mu$SR is widely applied to many kinds of materials these days. 

How to choose analysis functions is a key matter to deduce the information in physics from the $\mu$SR results in order to understand changes in electronic states in various temperature regions. For instance, local magnetic fields at the muon site in the paramagnetic state are well known to be coming from surrounding nuclear dipole moments, resulting in the formation of the Gaussian field distribution at the muon site. The Gaussian distribution typically occurs when there are independent contributions from many magnetic sources with similar amount of contribution. Simple metals like Cu are good examples to realize this situation. The local field observed at the muon site in Cu is produced by nuclear dipole moments of Cu surrounding the muon and satisfies the condition to realize the Gaussian distribution.\cite{Hayano,UemuraMuon} In this case, the $\mu$SR time spectrum is well described by the Gaussian Kubo-Toyabe (GKT) function, $P_{GKT}(t)$, as follows.\cite{KuboToyabe,Kubo1981} 
\begin{equation}
P_{GKT}(t) = \frac{1}{3} +\frac{2}{3} (1-\Delta^2t^2) \exp(-\frac{\Delta^2t^2}{2}).
\label{eq_GKT}
\end{equation}

\noindent
Here, $\Delta$ is the half width of the Gaussian distribution of magnetic field at the muon site. This GKT-type relaxation behavior of the $\mu$SR time spectrum is well observed in the paramagnetic state of many kinds of materials.\cite{Hayano,UemuraMuon,Kadono,IW5} 

The Lorentzian distribution tends to occur when contributions from one magnetic source dominates among others. One typical example is a dilute spin-glass system. In this case, one magnetic spin, which is located nearest to the muon, tends to give a dominant contribution.\cite{UemuraMuon} In those low-density spin systems, the local field due to the random and sparse magnetic dipole has an axial magnetic field distribution proportional to $B^2/(a^2 + \gamma_\mu^2 B^2)^2$ for the dilute limit (effectively for concentrations less than 3$\sim$5 \%), called Lorentzian-field\cite{Walstedt1974}. The $B$ is the distributed magnetic fields with the half width of $a$ and $\gamma_\mu$ is the muon's gyromagnetic ratio ($\gamma_\mu$ = 2$\pi \times$135.5 MHz/T). This situation is described by the Lorentzian Kubo-Toyabe (LKT) function, $P_{LKT}(t)$, which is as follows.\cite{UemuraMuon} 
\begin{equation}
P_{LKT}(t) = \frac{1}{3} +\frac{2}{3} (1-at) \exp(-at).
\label{eq_LKT}
\end{equation}

An intermediate $\mu$SR time spectrum can be considered as a cross-over distribution, which has characteristics somewhere between Gaussian and Lorentzian. For instance, in case that there are two independent field contributions, one having the Gaussian distribution and the other Lorentzian, the intermediate local field distribution can be realized. Another possible case is when the source is from one magnetic origin but the number of contributing magnetic spins is small though not one. 

Recently, another example to realize the intermediate $\mu$SR time spectrum was reported in the case that systems had non-uniform and/or low density distributions of nuclear magnetic moments. Organic molecular superconductors are typical examples showing this kind of distribution.\cite{Imajo,Kobayashi,Dita} Those systems have low dimensional crystal structures and low-density alignments of nuclear magnetic moments, resulting in the deformation of the Gaussian distribution of magnetic fields.

When a magnetic transition appears with decreasing temperatures, the intermediate $\mu$SR time spectrum is frequently observed around magnetic transition temperatures by a different reason from those mentioned above. Near the magnetic transition temperature, additional internal fields which are coming from surrounding fluctuating electronic magnetic moments appear at the muon site. Those additional fields are a couple of orders as large as those coming from nuclear magnetic moments. With approaching to the magnetic transition temperature, fluctuating electronic magnetic fields become mandatory and the $\mu$SR time spectrum changes from  Gaussian to Lorentzian reflecting the spin-spin correlation function.\cite{UemuraMuon,IW5,AdachiZnNi,Anand,IW6,Panagopo} Some examples showing this kind of changes in the time spectrum were reported in the La-based high $T_{\rm c}$ superconducting oxide, La$_{2-x}$Sr$_x$CuO$_4$, especially in the underdoped regime.\cite{IW6,Panagopo}

In those cases, phenomenological functions were used to analyze intermediate time spectra. One example is;
\begin{equation}
e^{-\lambda t} \times P_{GKT}(t).
\label{eq_LG}
\end{equation}

\noindent
Here, $\lambda$ is regarded as the dynamic relaxation rate of the muon-spin polarization which is caused by fluctuating electronic spins. This phenomenological function is used on the basis of the assumption that the measured system contained localized electronic moments which are fluctuating in time. However, those two parameters sometimes cause the trading-off effect to describe the intermediate time spectrum resulting in failures to reveal realistic electronic states. 

The Stretched Kubo-Toyabe (SKT) function is also well-used analysis function.\cite{Crook1997} 
\begin{equation}
P_{SKT}(t) = \frac{1}{3} + \frac{2}{3} (1-(\lambda t)^\alpha) \exp(-(\lambda t)^\alpha/\alpha)).
\label{eq_SKT}
\end{equation}

\noindent
Here, $\alpha$ ($\alpha$=1-2) and $\lambda$ are the stretch parameter and the relaxation rate of the muon-spin polarization, respectively. The $P_{SKT}(t)$ matches with $P_{LKT}(t)$ at $\alpha$=1 and $P_{GKT}(t)$ at $\alpha$=2. Although this form has been widely used for many $\mu$SR results because of the easiness of programing in the fit, it is difficult to get physical ideas how $\alpha$ and $\lambda$ can be related to the actual field distribution and the spin dynamics. 

Therefore, it is important to describe the intermediate muon-spin relaxation function in order to study crossover phenomena under the co-existence of two random and static (in the time-range of muon spin precession) magnetic fields which are independent to each other. Until now, various analysis functions have been developed to describe the $\mu$SR time spectrum, however, the description of the intermediate $\mu$SR time spectrum has not yet been successful enough. Phenomenological equations to mix the Gaussian and Lorentzian functions were examined,\cite{PSI,JSLord2005} and one generalized theoretical function was suggested for the analysis of the intermediate state.\cite{Kyoto} Those recent suggestions prove the high interest in developing the analysis function for the intermediate $\mu$SR time spectrum and its requirement is becoming higher year by year. 

For the current study, we described the crossover field in terms of a convoluted function of Gaussian and Lorentzian. We derived the equation of the three-dimensional (3D) convolution in two ways. The first derivation uses the convolution integral starting directly in the 3D space. The other derivation starts from that of the one-dimensional (1D) convolution and make it to be converted to the 3D form. From the latter, we showed that the equation can be decomposed to a sum of three known convolutions. By applying the Fourier transform to this equation, we achieved the correct relaxation function for the zero-field condition, which was found to be given by a simple analytical equation. In addition, we tried to describe the intermediate analysis function under applied magnetic fields and under dynamic fluctuations on the basis of the development of the zero-field intermediate analysis function. Finally, we applied our developed analysis function to some $\mu$SR results in order to make sure its validity. 

\section{Field distribution and relaxation function under coexistence of Gaussian and Lorentzian}

\begin{figure}
\begin{center}
\includegraphics[width=8cm]{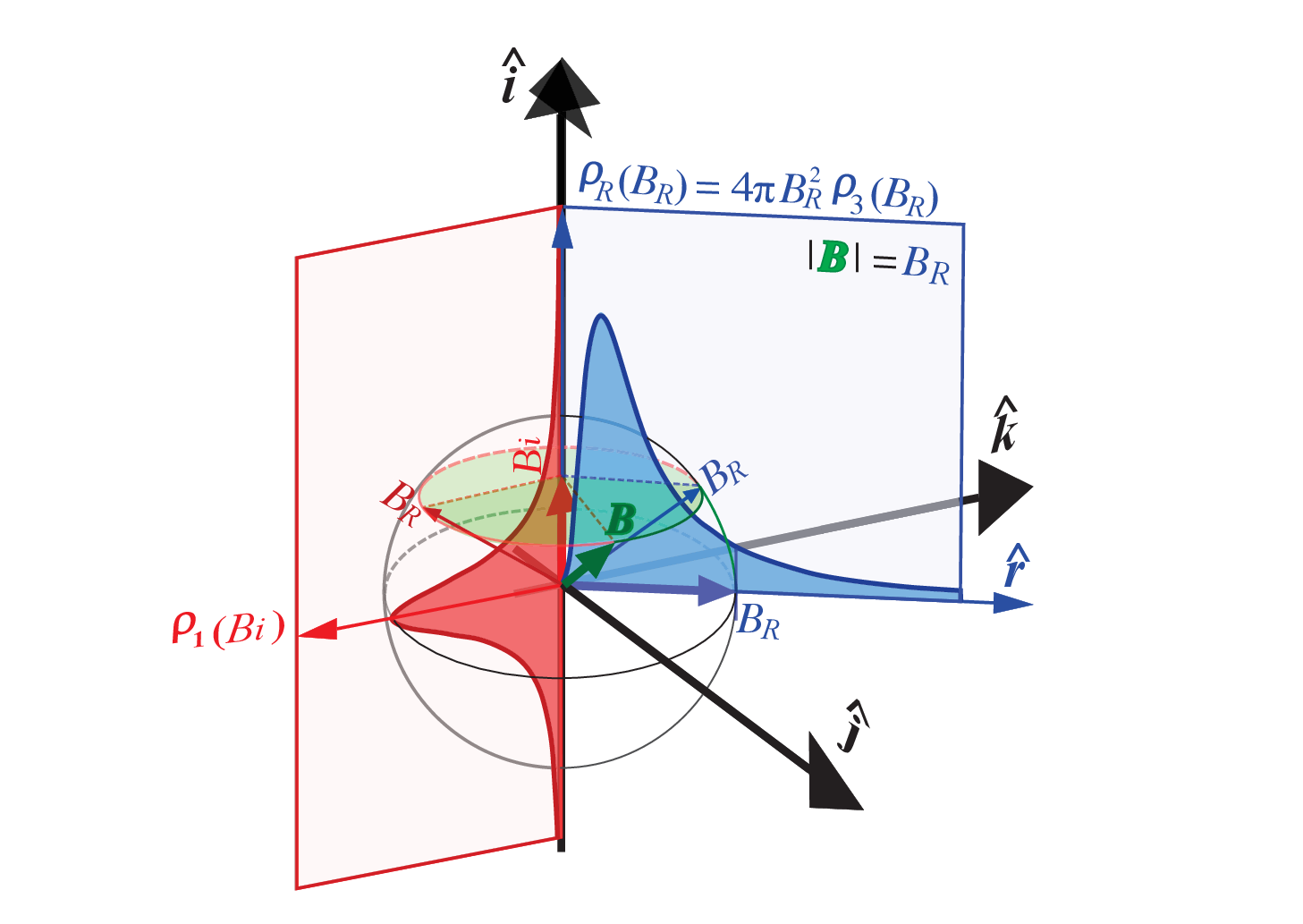}%
\end{center}
\caption{
Relation between the 3D field distribution, $\rho_3({\bm B})$, and its distribution projection in one dimension, $\rho_1(B_i)$. We also define the field size distribution, $\rho_R(B)$. They are normalized so that $\int_{-\infty}^{+\infty} \rho_1(B_i) dB_i = 1$, $4\pi \int_0^{+\infty} \rho_3(B) B^2 dB = 1$, and $\int_0^{+\infty} \rho_R(B) dB = 1$.}
\label{fig_1d_3d}
\end{figure}

\subsection{Conversion between 3D and 1D magnetic field distributions}

\ \ \ We start from showing how the 1D and 3D distributions of magnetic fields can be related when the field direction is random (namely, isotropic). First, we define the probability of finding a site with the magnetic field ${\bm B} = (B_x,B_y,B_z)$ as $\rho_3({\bm B})d^3{\bm B}$ (see Fig.\ref{fig_1d_3d}). If the field distribution is isotropic with $\rho_3({\bm B})$ having no dependence on the direction, we may write
$\rho_3({\bm B})d^3{\bm B} = \rho_3(B)B^2dBd(\cos{\theta})d\phi$, where $B$ is the size of the local field. We also define the distribution of the field size as $\rho_R(B)dB$, then $\rho_R(B) = 4{\pi}B^2 \rho_3(B)$. The distribution of the field component in one direction, for example $B_z$, is given by $\rho_1(B_z)dB_z$. In the cylindrical coordinate $(B_z,B_\rho,\phi)$, we get by projection 
\begin{equation}
\rho_1(B_z) = \int_0^{\infty} \rho_3(B) 2{\pi} B_{\rho} dB_{\rho} 
\end{equation}

\noindent
where $B^2=B_z^2+B_{\rho}^2$ and the integration is done keeping $B_z$ constant. Using $B dB = B_{\rho} dB_{\rho}$, 
\begin{equation}
\rho_1(B_z) = \int_{B_z}^{\infty} \rho_3(B) 2{\pi} B dB.
\end{equation}

\noindent
It follows, 
\begin{equation}
d\rho_1(B_z)/dB_z = -2{\pi} B_z \rho_3(B_z).
\end{equation}

\noindent
As the expression of the variable does not matter, we rewrite Eq.(5) as 
\begin{equation}
\rho_3(B) = -(1/2{\pi}B) d\rho_1(B)/dB
\label{eq_1to3}
\end{equation}

\noindent
and
\begin{equation}
\rho_R(B) = -2B d\rho_1(B)/dB.
\label{eq_1toR}
\end{equation}

\noindent
We set two distributions of $B$, Gaussian and Lorentzian. Each distribution is characterized by $\Delta$ or $a$ as the width of the distribution. For Gaussian, we get 
\begin{align}
\rho_{1,G}(B) & = (\gamma_\mu/\sqrt{2\pi}\Delta) \exp(-\gamma_\mu^2B^2/2\Delta^2),  \\
\rho_{3,G}(B) & = (\gamma_\mu^3/(2\pi)^{3/2} \Delta^3) \exp(-\gamma_\mu^2B^2/2\Delta^2),  \\
\rho_{R,G}(B) & = (2^{1/2}\gamma_\mu^3/\pi^{1/2} \Delta^3) B^2 \exp(-\gamma_\mu^2B^2/2\Delta^2).
\end{align}

\noindent
For the Lorentzian case, 
\begin{align}
\rho_{1,L}(B) & = (\gamma_\mu/\pi) a/(a^2+\gamma_\mu^2B^2), \\
\rho_{3,L}(B) & = (\gamma_\mu^3/\pi^2) a/(a^2+\gamma_\mu^2B^2)^2, \\
\rho_{R,L}(B) & = (4\gamma_\mu^3/\pi) aB^2/(a^2+\gamma_\mu^2B^2)^2. 
\end{align}

\subsection{3D convolution of the static magnetic field distribution}

\ \ \ First, we describe the distribution $\rho_{GL}(B)$ as the convolution of the Gaussian and Lorentzian fields. For the distribution of the summed field component in one direction, we can use 1D convolution, 
\begin{equation}
\begin{split}
\rho_{1,GL}(B) & = \int dB_1 \rho_{1,G}(B-B_1) \rho_{1,L}(B_1) \\
& = (a\gamma_\mu^2/\sqrt{2}\pi^{3/2}\Delta) \int dB_1 \exp(-\gamma_\mu^2(B-b_1)^2/2\Delta^2) / (a^2+\gamma_\mu^2B^2).
\label{eq_conv1d}
\end{split}
\end{equation}

\noindent
To obtain the 3D distribution $\rho_3({\bm B})$ of the vector summed field, we need 3D convolution. The convolution should be done by varying one of the fields, ${\bm B_1}$, while the other field is automatically determined. This leads to that ${\bm B_2} = {\bm B} - {\bm B_1}$. The probability of having ${\bm B_1}$ and ${\bm B_2}$ at the same time is 
$\rho_{3,G}({\bm B_2}) \rho_{3,L}({\bm B_1})$. This probability should be integrated for all possible ${\bm B_1}$ to find the probability of having field ${\bm B}$. Thus, 
\begin{equation}
\begin{split}
\rho_{3,GL}({\bm B}) 
& = \int d^3{\bm B_1} \rho_{3,G}({\bm B_2}) \rho_{3,L}({\bm B_1}) \\
& = 2\pi \int \int B_1^2 dB_1 d(\cos{\theta_1}) \rho_{3,G}(B_2) \rho_{3,L}(B_1)\\
& = (\gamma_\mu^3/(2\pi)^{3/2}\Delta^3) (a\gamma_\mu^3/\pi^2) \int \int dB_1d(\cos{\theta_1}) \exp(-\gamma_\mu^2B_2^2/2\Delta^2) B_1^2/(a^2+\gamma_\mu^2B_1^2)^2.
\end{split}
\end{equation}

\noindent
where $B_2^2 = B^2 + B_1^2 -2B_1B \cos{\theta_1}$. The integration over $\cos{\theta_1}$ can be done analytically and this gives
\begin{equation}
\begin{split}
\rho_{3,GL}(B) 
& = (\sqrt{2}a\gamma_\mu^4/\pi^{3/2}\Delta) \int_0^{+\infty} dB_1B B_1 \\
& \times  [\exp(-\gamma_\mu^2(B-B_1)^2/2\Delta^2) 
- \exp(-\gamma_\mu^2(B+B_1)^2/2\Delta^2)] 
/(a^2+\gamma_\mu^2B_1^2)^2 \\
& =  (\sqrt{2}a\gamma_\mu^4/\pi^{3/2}\Delta) \int_{-\infty}^{+\infty} dB_1 B B_1
\exp(-\gamma_\mu^2(B-B_1)/2\Delta^2)/(a^2+\gamma_\mu^2B_1^2)^2.
\label{eq_conv3d}
\end{split}
\end{equation}

Next, we show another derivation of the 3D convolution form. When two independent distributions contribute, the projected sum of fields is represented by the 1D convolution, 
\begin{equation}
\rho_{1,GL}(B) = \int_{-\infty}^{+\infty} dB_1 \rho_{1,G}(B-B_1) \rho_{1,L}(B_1).
\end{equation}

\noindent
As the sum of field distribution is also isotropic, using the relation Eq.(\ref{eq_1toR}), we get
\begin{equation}
\begin{split}
\rho_{R,GL}(B) & = -2B d\rho_{1,GL}(B)/dB \\
& = \int dB_1 (-2B)(d\rho_{1,G}(B-B_1)/dB) \rho_{1,L}(B_1) 
\end{split}
\end{equation}

\noindent
It can be shown that this lead to the same form as Eq.(\ref{eq_conv3d}). 
However, instead, we here derive another form, applying the Fourier transform to obtain the relaxation function.
\begin{multline}
\rho_{R,GL}(B) = \int dB_1 (-2(B-B_1)-2B_1) 
(d\rho_{1,G}(B-B_1)/dB) \rho_{1,L}(B_1) 
\end{multline}
from the relations, $-2(B-B_1)d\rho_{1,G}(B-B_1)/dB = \rho_{R,G}(B-B_1)$ and 
$d\rho_{1,G}(B-B_1)/dB = -d\rho_{1,G}(B-B_1)/dB_1$, 
\begin{equation}
\begin{split}
\rho_{R,GL}(B) & =  \int_{-\infty}^{+\infty} dB_1 \rho_{R,G}(B-B_1) \rho_{1,L}(B_1) 
+ \int_{-\infty}^{+\infty} dB_1 (d\rho_{1,G}(B-B_1)/dB_1) 2B_1 \rho_{1,L}(B_1) \\
& =  \int_{-\infty}^{+\infty} dB_1 \rho_{R,G}(B-B_1) \rho_{1,L}(B_1) 
+ [\rho_{1,G}(B-B_1) 2B_1 \rho_{1,L}(B_1)]_{-\infty}^{+\infty} \\
& - \int_{-\infty}^{+\infty} dB_1 \rho_{1,G}(B-B_1) d(2B_1 \rho_{1,L}(B_1))/dB_1 \\
& =  \int_{-\infty}^{+\infty} dB_1 \rho_{R,G}(B-B_1) \rho_{1,L}(B_1) 
- \int_{-\infty}^{+\infty} dB_1  \rho_{1,G}(B-B_1) (2B_1 d\rho_{1,L}(B_1)/dB_1) \\
& - \int_{-\infty}^{+\infty} dB_1  \rho_{1,G}(B-B_1) 2\rho_{1,L}(B_1) \\
& =  \int_{-\infty}^{+\infty} dB_1 \rho_{R,G}(B-B_1) \rho_{1,L}(B_1)  
+  \int_{-\infty}^{+\infty} dB_1 \rho_{1,G}(B-B_1) \rho_{R,L}(B_1)  \\
& -  2 \int_{-\infty}^{+\infty} dB_1 \rho_{1,G}(B-B_1) \rho_{1,L}(B_1).
\label{eq_3conv}
\end{split}
\end{equation}

\noindent
As the above handling is purely mathematical, we should note that $\rho_R$'s are defined even in negative $B$ range by Eq.(\ref{eq_1toR}) and $\rho_R(-B) = \rho_R(B)$ as $\rho_1(B)$'s are assumed symmetric.

\begin{figure}
\begin{center}
\includegraphics[width=8cm]{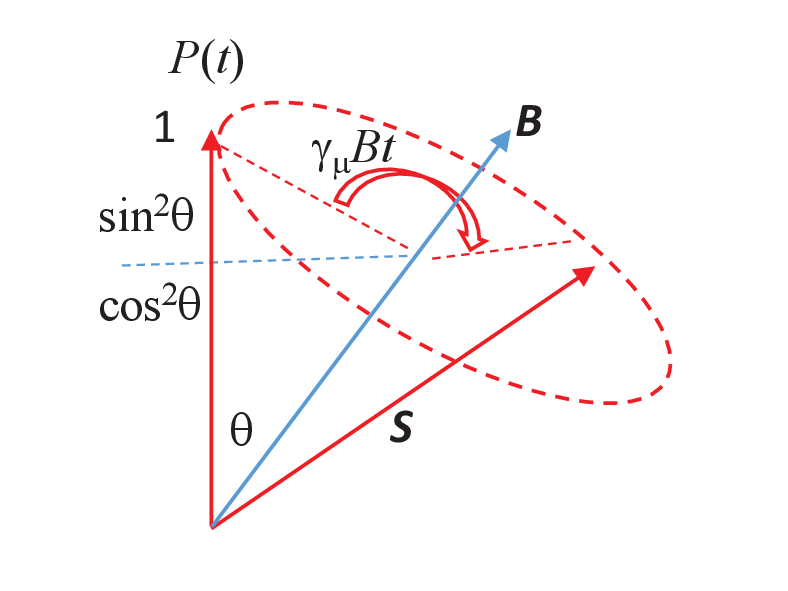}%
\end{center}
\caption{\label{fig_precession}
The spin rotation in zero-field. Decomposition of the muon polarization into the spin-conserving part, $\cos^2 \theta$, and the spin-precession part, $\sin^2 \theta$, are given. The precession part rotates around the internal magnetic field ${\bm B}$ with the Larmor frequency of $\gamma_{\mu}B$.}
\end{figure}

\subsection{Muon spin relaxation function under isotropic field distribution}

\ \ \ Now, let's discuss muon spin in referring to internal-field distribution $\rho_R(B)$. Hence, muons have a polarization-axis as ensemble, and the polarization can be depolarized (relaxed) in time due to the spin-precession around the internal field, because each muon will sense a different magnetic field at the specific position of the muon. For simplicity, let's describe in a semi-classical manner. An example of muon-spin-precession is schematically illustrated in Fig. \ref{fig_precession}. Taking the quantum axis to be in the direction of the muon polarization at $t=0$, $\theta$ and $\phi$ are the polar and azimuthal angles of $B$ at the muon site, respectively. The $B$ distributes randomly in the angle referring to the quantum-axis. Its field strength is described by the 3D distribution $\rho_3(B)$ (or the size distribution $\rho_R(B)$). The muon spin precesses around $B$ with the Larmor precession frequency, $\omega_{\mu}$, where $\omega_{\mu}$=$\gamma_{\mu}B$. By taking ensemble, the component vertical to the initial polarization is canceled out because of the symmetry and only the spin polarization parallel to the initial spin remains,  
\begin{equation}
\begin{split}
P(t) 
& = \int \int \int d^3{\bm B} 
[\cos^2 \theta +  \sin^2 \theta \cos(\gamma_{\mu}Bt)] \rho_3(B) \\
& = \int dB \int d(\cos{\theta})
[\cos^2 \theta +  \sin^2 \theta \cos(\gamma_{\mu}Bt)] \frac{1}{2} \rho_R(B). 
\end{split}
\end{equation}

In the case of an isotropic field distribution, $\rho_R(B)$ is independent of $\theta$, so the we can take an integral over $\cos{\theta}$, resulting in 
\begin{equation}
P(t) = \frac{1}{3} + \frac{2}{3} \int_0^{\infty} \cos(\gamma_{\mu}Bt) \rho_R(B) dB
= \frac{1}{3} + \frac{2}{3} P_{osc}(t)
\end{equation}

\noindent
The $P_{osc}(t)$ is the oscillation component of the muon-spin relaxation. We here define two Fourier transform, one in the range  0 to $+\infty$ and the other  in the range $-\infty$ to $+\infty$ as follows, 
\begin{align}
&\hat{\rho}^+(t) = \int_0^{+\infty} \cos(\gamma_{\mu}Bt) \rho(B) dB, \nonumber\\
&\hat{\rho}(t) = \int_{-\infty}^{+\infty} \cos(\gamma_{\mu}Bt) \rho(B) dB.
\end{align}

\noindent
For symmetric distribution $\rho(B)$, $\hat{\rho}(t) =  2\hat{\rho}^+(t)$.

Now, we set $P_{osc}(t) = \hat{\rho}_{R,GL}^+(t)$, where the field distribution is given as the convolution Eq.(\ref{eq_3conv}). 
\begin{equation}
\begin{split}
\hat{\rho}^+_{R,GL}(t) 
& = \frac{1}{2}\hat{\rho}_{R,GL}(t) \\
& = \frac{1}{2} \int_{-\infty}^{\infty} \cos(\gamma_{\mu}Bt) \rho_{R,GL}(B) dB \\
& = \frac{1}{2}  \int_{-\infty}^{\infty} \cos(\gamma_{\mu}Bt) 
[\int_{-\infty}^{+\infty} dB_1 \rho_{R,G}(B-B_1) \rho_{1,L}(B_1)]  dB \\
& +  \frac{1}{2} \int_{-\infty}^{\infty} \cos(\gamma_{\mu}Bt) 
[\int_{-\infty}^{+\infty} dB_1 \rho_{1,G}(B-B_1) \rho_{R,L}(B_1)] dB \\
& - \int_{-\infty}^{\infty} \cos(\gamma_{\mu}Bt) 
[\int_{-\infty}^{+\infty} dB_1  \rho_{1,G}(B-B_1) \rho_{1,L}(B_1)] dB
\end{split}
\end{equation}

\noindent
Using the well-known principles of the Fourier transform of functions $f$ and $g$, additive principle ($\widehat{f+g} = \hat{f} + \hat{g}$) and convolution principle ($\widehat{f*g} = \hat{f} \times \hat{g}$), where $f*g$ means convolution, 
\begin{equation}
\begin{split}
\hat{\rho}_{R,GL}^+(t) 
& = \frac{1}{2}  \int_{-\infty}^{\infty} \cos(\gamma_{\mu}Bt) \rho_{R,G}(B) dB 
\int_{-\infty}^{+\infty} \cos(\gamma_{\mu}Bt) \rho_{1,L}(B)  dB \\
& + \frac{1}{2}  \int_{-\infty}^{\infty} \cos(\gamma_{\mu}Bt) \rho_{1,G}(B) dB 
\int_{-\infty}^{+\infty} \cos(\gamma_{\mu}Bt) \rho_{R,L}(B)  dB \\
& - \int_{-\infty}^{\infty} \cos(\gamma_{\mu}Bt) \rho_{1,G}(B) dB 
\int_{-\infty}^{+\infty} \cos(\gamma_{\mu}Bt) \rho_{1,L}(B)  dB \\
& = \frac{1}{2}\hat{\rho}_{R,G}(t)\hat{\rho}_{1,L}(t) + \frac{1}{2}\hat{\rho}_{1,G}(t)\hat{\rho}_{R,L}(t) - \hat{\rho}_{1,G}(t)\hat{\rho}_{1,L}(t) \\
& = \hat{\rho}_{R,G}^+(t)\hat{\rho}_{1,L}(t) + \hat{\rho}_{1,G}(t)\hat{\rho}_{R,L}^+(t) - \hat{\rho}_{1,G}(t)\hat{\rho}_{1,L}(t)
\end{split}
\end{equation}

Note that the relation is applicable as far as the two distributions are independent and both isotropic. In a special case when the two distributions are Gaussian and Lorentzian, their Fourier counterparts are well-known including those for 3D distributions,\cite{KuboToyabe, Hayano, Kubo1981}
\begin{align}
&\hat{\rho}_{1,G}(t) = \exp(-\Delta^2t^2/2) \\
&\hat{\rho}_{R,G}^+(t) = (1-\Delta^2t^2) \exp(-\Delta^2t^2/2) \\
&\hat{\rho}_{1,L}(t) = \exp(-at) \\
&\hat{\rho}_{R,L}^+(t) = (1-at) \exp(-at).
\end{align}

\noindent
We get
\begin{equation}
\hat{\rho}_{R,GL}^+(t) \  =  (1-\Delta^2t^2-at) \exp \left(-\Delta^2t^2/2 \right) \exp(-at) 
\end{equation}

\noindent
as the oscillation part. Thus, the relaxation function under random directional field distribution is 
\begin{equation}
P_{GLKT}(t) = \frac{1}{3} +\frac{2}{3} (1-\Delta^2t^2-at) \exp \left(-\frac{\Delta^2t^2}{2}-at \right) 
\label{eq_GLKT}
\end{equation}

\noindent
This is the correct extension form of the Kubo-Toyabe relaxation function \cite{KuboToyabe,Kubo1981} for the convoluted distribution of Gaussian and Lorentzian. The function becomes Gaussian Kubo-Toyabe if $a$=0 and Lorentzian Kubo-Toyabe if $\Delta$=0. The same function was mentioned in \cite{PSI,JSLord2005} although no detail derivations were shown there.

\begin{figure}
\begin{center}
\includegraphics[width=8cm]{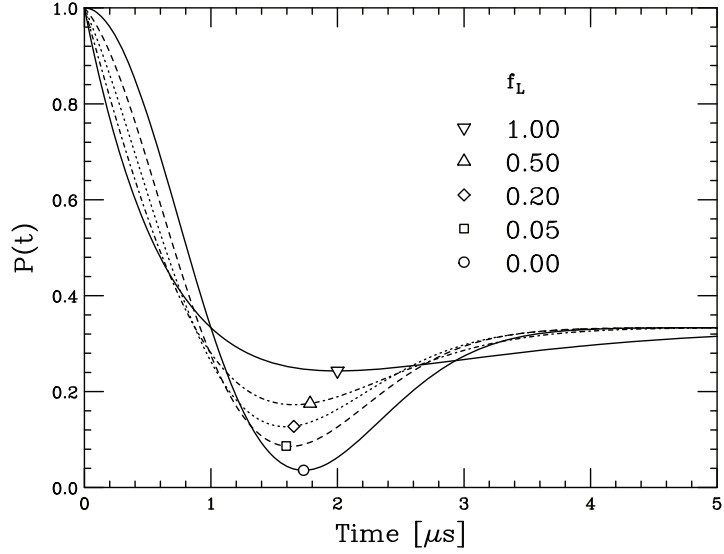}%
\end{center}
\caption{
Simulation of Eq.(\ref{eq_GLKT}) with different mixing ratio of Lorentzian and Gaussian, $f_L$ = 0, 0.05, 0.2, 0.5 and 1, while keeping $\sqrt{\Delta^2+a^2}$ = 1 $\mu s^{-1}$.
The locations of the minimum are also shown by open marks. }
\label{fig_GLKT}
\end{figure}

The behavior of Eq.(\ref{eq_GLKT}) is graphically shown in Fig. \ref{fig_GLKT} by changing the fraction of Lorentzian source contribution $f_L = a^2/(\Delta^2+a^2)$ while keeping $\sqrt{\Delta^2+a^2}$ = 1 $\mu s^{-1}$. One of the most characteristic features of the relaxation function is the dip. The location of the minimum of the dip can be found by taking the derivative of $P_{GLKT}(t)$ and solving the cubic equation

\begin{equation}
\Delta^4t^3+2a\Delta^2t^2+(a^2-3\Delta^2)t^2-2a = 0.
\end{equation}

\noindent
Using Cardano's method, we get as the solution 

\begin{equation}
t_{min} = \frac{2}{3} [\sqrt{b^2+9}\cos(\phi/3) - b]/\Delta,
\end{equation}

\noindent
where $b = a/\Delta$, and $\phi$ $(0\sim\frac{\pi}{2})$ is chosen so that $\tan{\phi} = \sqrt{(1+9/b^2)^3-1}$. In here, the Gaussian and Lorentzian distributions have the minimum dip in their shape at $t_{min} = \sqrt{3}/\Delta$ and $2/a$, respectively.

\begin{figure}
\begin{center}
\includegraphics[width=8cm]{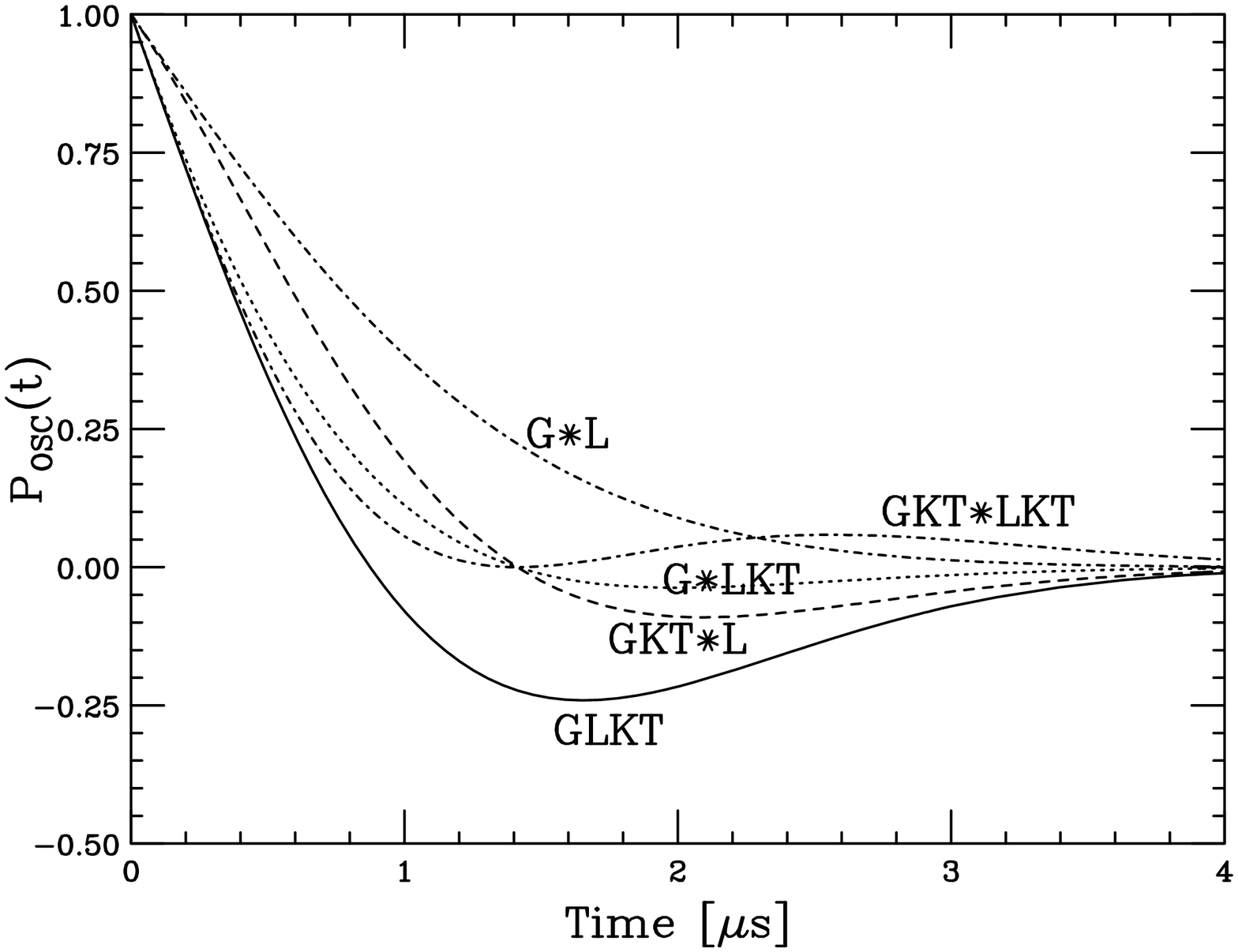}%
\end{center}
\caption{
Comparison of the oscillating part of the relaxation function of Eq.(\ref{eq_GLKT}) with other functions given as products of the relaxation functions of Gaussian- and Lorentzian-origin with $\Delta = 0.707 {\mu}s^{-1}$ and $a = 0.707 {\mu}s^{-1}$, respectively. G : Gaussian, GKT: Gaussian Kubo-Toyabe, L: Lorentzian, LKT: Lorentzian Kubo-Toyabe, and GLKT: extended Kubo-Toyabe function [Eq.(\ref{eq_GLKT})] for the convolution of Gaussian and Lorentzian. 
G*L: $\exp(-\frac{\Delta^2t^2}{2}-at)$ ,
GKT*L: $(1-\Delta^2t^2) \exp(-\frac{\Delta^2t^2}{2}-at) $,
G*LKT: $(1-at) \exp(-\frac{\Delta^2t^2}{2}-at) $,
GKT*LKT: $(1-\Delta^2t^2) (1-at) \exp(-\frac{\Delta^2t^2}{2}-at)$ , and
GLKT: $(1-\Delta^2t^2-at) \exp(-\frac{\Delta^2t^2}{2}-at) $.
}
\label{fig_products}
\end{figure}

\section{Comparison with other relaxation functions}

\ \ \ There have been used several different relaxation functions in an attempt to fit the $\mu$SR time spectrum in the cross-over regime. Typical trials were to approximate the relaxation as a product of functions of Gaussian and Lorentzian origin. The dip described in Eq.(\ref{eq_GLKT}) can be compared with several different combinations of the product in Fig. \ref{fig_products}. Unfortunately, it is obvious that no other function form is successful in reproducing the correct form. 
 
We tested how the SKT function, Eq.(\ref{eq_SKT}), can be compared to the exact form. Since there is no equation known relating $\alpha$ and $\lambda$ to $\Delta$ and $a$, $\alpha$ and $\lambda$ were just chosen, making the functions the best matched. Figure 5 shows a reasonable match as seen for the case of $f_L = 0.5$. Table \ref{tab_alpha} shows the fitted $\alpha$ and $\lambda$ parameters for several mixing ratios. The root-mean-square (RMS) deviation from the exact function is also shown. The stretched function parameters seem to reasonably approximate the exact function within the RMS deviation $\sim$1\%. However, some differences are evident such as the slower decrease in $P_{SKT}(t)$ at time zero. Note that the physics basis of the SKT function is vague compared to the exact form. 

\begin{figure}
\begin{center}
\includegraphics[width=8cm]{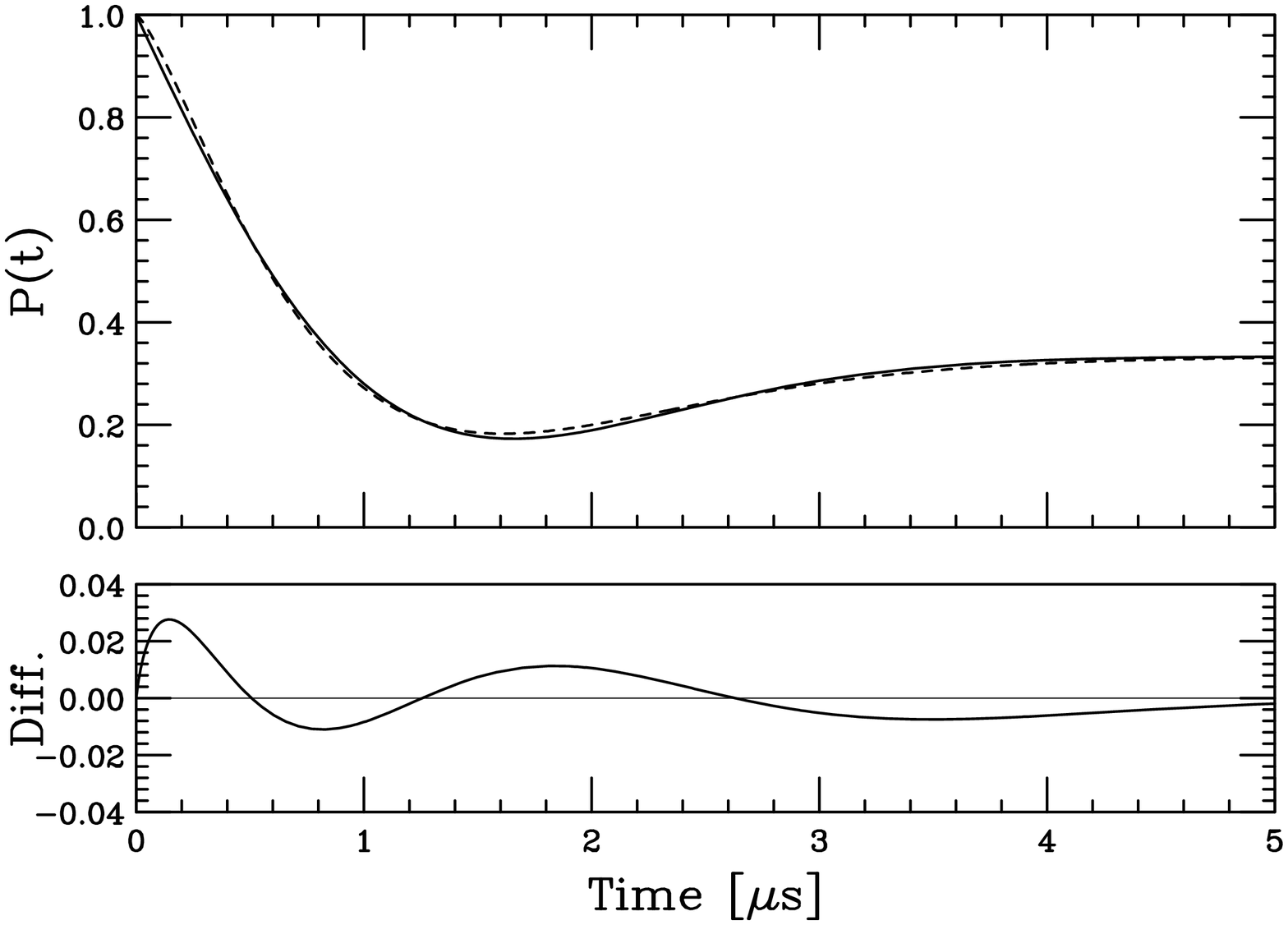}
\end{center}
\caption{
Top: Comparison of the exact relaxation function Eq.(\ref{eq_GLKT}) (solid line), and Eq.(\ref{eq_SKT}) (dashed line) for $f_L = 0.5$ with $\sqrt{\Delta^2 + a^2}$ = 1 $\mu s^{-1}$. Bottom: Difference between Eq.(\ref{eq_SKT}) and Eq.(\ref{eq_GLKT}). The best fit was made with $\alpha$ = 1.315 and $\lambda$ = 1.175 $\mu s^{-1}$.}
\label{fig_SKT_ZF}
\end{figure}

\begin{table}
\label{tab_alpha}
\caption{
Parameters of Eq.(\ref{eq_SKT}) given by a fit to Eq.(\ref{eq_GLKT}). The $f_L$ is the mixing ratio of the Lorentzian source. The RMS is the root-mean-square discrepancy between the two functions.}
\begin{center}
\begin{tabular}{|ccc|ccc| } \hline
\multicolumn{3} {|c|} {Source distribution, $P_{GLKT} (t)$} & \multicolumn{3} {c|} {Stretched Kubo-Toyabe, $P_{SKT} (t)$} \\
$f_L$  &  $\Delta$   &      $a$      &   $\alpha_S$   &  $\lambda_S$  &  RMS \\
\hline
    0.00&         1.0000&         0.0000&    2.0000&    1.0000&     0.0000 \\
    0.25&         0.8660&         0.5000&    1.4933&    1.1791&     0.0083 \\
    0.50&         0.7071&         0.7071&    1.3146&    1.1748&     0.0080 \\
    0.75&         0.5000&         0.8660&    1.1638&    1.1178&     0.0059 \\
    1.00&         0.0000&         1.0000&    1.0000&    1.0000&     0.0000 \\ 
\hline
\end{tabular}
\end{center}
\end{table}

\section{Responses of the intermediate analysis function against external parameters}
\subsection{Responses to magnetic fields}

\ \ \ The $\mu$SR experiment in the zero-field condition is the unique and strong advantage to use the muon which has the self-polarization along of its initial spin direction. In addition to this, responses of the $\mu$SR time spectrum in magnetic fields applied from outside to materials are also important to investigate dynamic and static properties of local fields at the muon site.\cite{Hayano,UemuraMuon} In order to investigate dynamic properties of local fields at the muon site, the magnetic field is applied along the same direction of the initial muon-spin polarization. We call this applied magnetic field as the longitudinal field (LF). Accordingly, we also created the general formation to describe the magnetic field dependence of our developed intermediate analysis function.  

In order to describe the LF dependence of the $\mu$SR time spectrum, we need to add is LF with the amount of $B_{\rm 0}$ along the quantum axis which is the same with the initial muon-spin polarization. Since it was not so easy to write down the LF dependence following the same detail manner from the concept drawn in Fig. 1, we used a different way to derive the final equation. That is to use the Kubo formula with the Fourier transform of the field distribution.\cite{Kubo1981}  
\begin{equation}
	P_z(t,B_0) = 1-2t\left(\frac{d}{dt}[Q(t)]\right)\frac{\cos\omega_0 t}{(\omega_0 t)^2}+\frac{2}{\omega_0^2}\lim_{t \to 0}\left(\frac{\frac{d}{dt}[Q(t)]}{t}\right) + 2\int_0^t\frac{\sin\omega_0\tau}{\omega_0^3\tau}\frac{d}{d\tau}\left(\frac{\frac{d}{d\tau}[Q(\tau)]}{\tau}\right)\,d\tau. 
\label{eq_Kubo}
\end{equation}

Here, $\omega_{0}=\gamma_{\mu}B_{0}$. The $Q\left(t\right)$ is a Fourier transform of the convoluted distribution between Gaussian and Lorentzian. Referencing Eq.(\ref{eq_GLKT}),  $Q\left(t\right)$  is given as follows. 
\begin{equation}
    Q\left(t\right) = {\rm exp}\,(-at-\frac{\Delta^{2}t^{2}}{2}).
\label{eq_Qt}
\end{equation}

\noindent
Simply calculate this equation, we reach to the required equation to draw the LF dependence of the muon-spin polarization, $P_{LFGLKT}(t,B_{\rm 0})$, as; 
\begin{equation}
	\begin{split}
		P_{LFGLKT}\,(t,B_{\rm 0}) ={}&1-\frac{a}{\omega_{0}}\left(J_{1}\left(\omega_{0}t\right)\exp\left(-at-\frac{\Delta^{2}t^{2}}{2}\right)\right)\\&-\frac{2\Delta^{2}}{\omega_{0}^{2}}\left(1-\exp\left(-at-\frac{\Delta^{2}t^{2}}{2}\right)\cos\left(\omega_{0}t\right)\right)\\&-\frac{a^{2}}{\omega_{0}^{2}}\left(J_{0}\left(\omega_{0}t\right)\exp\left(-at-\frac{\Delta^{2}t^{2}}{2}\right)-1\right)\\&-\left[1+\left(\frac{a^{2}-3\Delta^{2}}{\omega_{0}^{2}}\right)\right]a\int_{0}^{t}J_{0}\left(\omega_{0}\tau\right)\exp\left(-a\tau-\frac{\Delta^{2}\tau^{2}}{2}\right)\,d\tau\\&-\left(\frac{a^{2}\Delta^{2}}{\omega_{0}^{2}}-\frac{2\Delta^{4}}{\omega_{0}^{3}}\right)\int_{0}^{t}\sin\left(\omega_{0}\tau\right)\exp\left(-a\tau-\frac{\Delta^{2}\tau^{2}}{2}\right)\,d\tau\\&-\frac{a\Delta^{2}}{\omega_{0}^{2}}\int_{0}^{t}\cos\left(\omega_{0}\tau\right)\exp\left(-a\tau-\frac{\Delta^{2}\tau^{2}}{2}\right)\,d\tau.
	\end{split}
\label{eq_LF_GLKT}
\end{equation}

\noindent
In here, $J_0$ and $J_1$ are the 0$^{th}$ and 1$^{st}$ order spherical Bessel functions, respectively. Other expressions are the same with those used in the previous sections. Figure 6 shows the schematic drawing of Eq.(\ref{eq_LF_GLKT}) in the case of $\Delta$ = $a$ = 0.707 $\mu$sec$^{-1}$ with changing LF. 

\begin{figure}[h]
\begin{center}
\includegraphics[width=10cm]{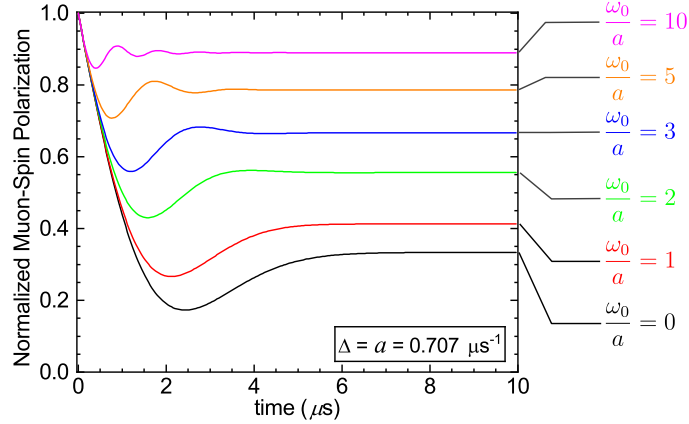}
\end{center}
\caption{Eq.(\ref{eq_GLKT}) against variable longitudinal fields with $\Delta$ = $a$ = 0.707 $\mu$sec$^{-1}$. In the case of $\frac{\omega_0}{a} \gg$ 10, the time spectrum is nearly decoupled from the internal field which has the intermediate distribution between Gaussian and Lorentzian. }
\end{figure}

The LF dependence of the time spectrum with increasing LF described by Eq.(\ref{eq_LF_GLKT}) has the similar characters to those of the Gaussian and Lorentzian functions.\cite{Hayano,UemuraMuon} Those are $i$) the dip in the time spectrum becomes smaller, $ii$) the so-called $\frac{1}{3}$-tail of the spectrum goes up and $iii$) the additional muon-spin precession around LF with the small amplitude appears in the earlier time region, and $iv$) the time spectrum becomes almost decoupled from local fields and locked along by LF keeping the initial muon-spin polarization in the case of $\frac{\omega_0}{a} \gg$ 10. 

\subsection{Responses to dynamic local fields}
\ \ \ In many cases, we need to discuss dynamic effects on the $\mu$SR time spectrum. Changes in local fields at the muon site in time are caused by magnetic transitions,\cite{UemuraLCO,Budnick,IW1,IW2,IW3,Zheng,O2,Pattenden,Aidy} molecular dynamics,\cite{Blundell}  ion/spin diffusions and muon motions.\cite{Nagamine,Ishida,Kadono,IW4,Sugiyama,Martin,PrattDif} If those dynamic changes in local fields happen within the $\mu$SR time window (10$^{-6}$-10$^{-11}$ sec), the $\mu$SR time spectrum is affected and shows different behavior from the static scenario which was given in previous sections. 

Accordingly, we describe the dynamic effect on the basis of Eq.(\ref{eq_GLKT}). In order to do this, we need to set some assumptions on the dynamic effect following the well established ways to take into account the dynamic motion of the muon.\cite{UemuraMuon} Those are $i$) local fields at the muon site do not change in time, $ii$) the muon is hopping in local fields, $iii$) the muon's motion can be described as the Markov process with the hopping frequency of $\nu$ on the basis of the strong-collision model, $iv$) the hopping frequency is within the $\mu$SR characteristic time window. 

What happen on the muon in those dynamic conditions is as follows. When the muon is trapped at one position at time $t$, the muon sees static local fields distributed at the muon position and shows the Larmor precession motion. The muon does not hop during a short time $t'$ after $t$ and depolarizes its spin polarization following Eq.(\ref{eq_GLKT}). Just after the muon hops to a next place after $t'$, the muon starts to see different local fields and depolarizes again around those different local fields following Eq.(\ref{eq_GLKT}) with the different initial condition from that given at $t$. After the hopping process is repeated within the $\mu$SR observation time which is typically up to around 20 $\mu$sec in the case of the use of a pulsed muon,\cite{NagamineMuon} the final $\mu$SR time spectrum, $P_{DGLKT}(t,\nu)$, can be described as the total sum of those hopping procedure as follows; 
\begin{multline}
	P_{DGLKT}\left(t,\nu\right) =
		\exp \left( -\nu t\right) \bigg[ P_{GLKT} \left( t\right)+\nu \int_{0}^{t} P_{GLKT}\left(t-t_{1}\right)P_{GLKT}\left(t_{1}\right) \,dt_{1}\\
		+\nu^{2}\int_{t_{2}}^{t}\int_{0}^{t_{2}}P_{GLKT}\left(t-t_{2}\right)P_{GLKT}\left(t_{2}-t_{1}\right)P_{GLKT}\left(t_{1}\right) \,dt_{2} \,dt_{1}+\cdots \bigg] 
\label{Dynaic_GLKT}	
\end{multline}

\noindent
The Eq.(\ref{Dynaic_GLKT}) is summarized as follows. 
\begin{equation}
	P_{DGLKT}\left(t,\nu\right)=\exp\left(-\nu t \right)P_{GLKT}\left(t\right) +\nu \int_{0}^{t}\exp\left(-\nu \left(t-t^\prime \right) \right)P_{GLKT}\left(t-t^\prime\right)P_{DGLKT}\left(t^\prime,\nu  \right)\,dt^\prime
\label{Dynamic_GLKT_Final}
\end{equation}

\noindent
Here, exp(-$\nu (t-t'))$ is the correlation function of the muon's hopping motion on the basis of the strongly collision model.\cite{UemuraMuon} The inverse of $\nu$ is related to the dynamic muon-spin depolarization rate. This equation has to be solved self-consistently because the right hand term includes the same depolarization term. 

\begin{figure}[h]
\begin{center}
\includegraphics[width=15cm]{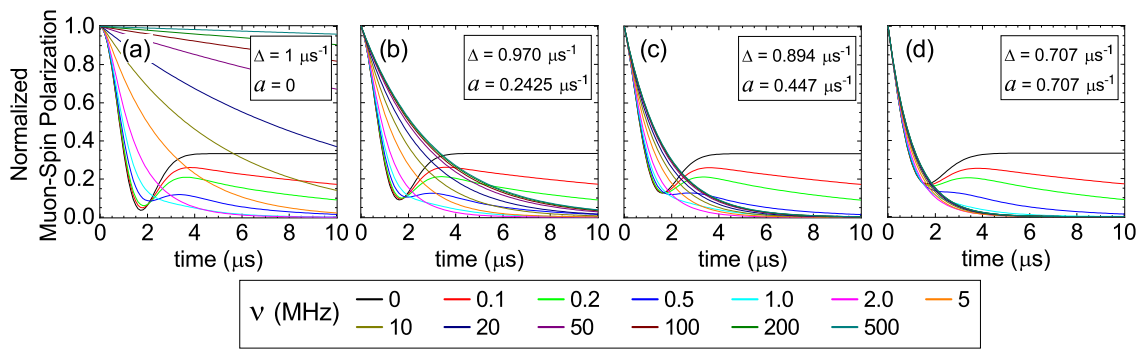}
\end{center}
\caption{Schematic picture of Eq.(\ref{Dynamic_GLKT_Final}) against variable $\nu$ to be 0, 0.2, 0.5, 1, 2, 5, 10, 20, 50, 100, 200 and 500 MHz. (a) The pure Gaussian case,\cite{UemuraMuon} (b) $\frac{a}{\Delta}$ = $\frac{1}{4}$, (c) $\frac{a}{\Delta}$ = $\frac{1}{2}$ and (d) $\frac{a}{\Delta}$ = 1. In the case of $\nu \gg$ 1 MHz, the time spectrum no longer has the so-called $\frac{1}{3}$-tail.  }
\end{figure}

Figure 7 shows a schematic picture of Eq.(\ref{Dynamic_GLKT_Final}) simulated by changing $\nu$ to be 0, 0.2, 0.5, 1, 2, 5, 10, 20, 50, 100, 200 and 500 MHz. $\Delta$ and $a$ were set to be some convenient values in order to make simulated time spectra to be easy to see within the experimental time region of $\mu$SR up to around 10 $\mu$sec.  
The overall picture of the response of the $\mu$SR time spectrum is different from that described by Eq.(\ref{eq_GKT}) (Fig. 7(a) \cite{UemuraMuon}) especially when the ratio $\frac{a}{\Delta}$ becomes large. The $\frac{1}{3}$-tail starts to relax first when the value of $\nu$ increases from the zero value. With increasing the value of $\nu$, the dip disappears and the $\frac{1}{3}$-tail can no longer be observed. The time spectrum tends to show no motional narrowing effect for the higher values of $\frac{a}{\Delta}$. This is because of the non-negligible LKT component in Eq.(\ref{Dynamic_GLKT_Final}) which is well known not to show the motional narrowing effect.\cite{Fiory,Leon,Silsbee}

\section{Comparison with $\mu$SR data}
\subsection{Muon-spin depolarization by distributed static local fields}

\ \ \ Candidate materials to which Eq.(\ref {eq_GLKT}) may be applied are organic molecules, especially organic molecular superconductors. The general tendency of the crystal structure of those kinds of organic systems shows low-dimensional and anisotropic states. In addition, atomic components of those organic systems contain only light elements that do not have large natural abundance of nuclear magnetic moments, like C and O. Those conditions can realize non uniform and dilute spin conditions.  

As an example, the intermediate $\mu$SR time spectrum was reported in the paramagnetic state of the low dimensional organic superconductor, $\lambda$-(BETS)$_2$GaCl$_4$ (BETS=(CH$_2$)$_2$S$_2$Se$_2$C$_6$Se$_2$S$_2$(CH$_2$)$_2$).\cite{Dita} The $\lambda$-(BETS)$_2$GaCl$_4$ shows the superconducting state below about 5.3 K and does not have any clear localized magnetic moment.\cite{Imajo,Kobayashi,Dita} $\mu$SR time spectrum showed the intermediate shape and was independent of temperature in the paramagnetic state.\cite{Dita} We can technically analyze this intermediate $\mu$SR time spectrum by using Eq.(\ref{eq_LG}). However, this method is hard to be appropriate because almost no localized electronic magnetic moment is expected in this system. From the view point of $\mu$SR, the nuclear dipole field is well recognized to be time independent due to the higher frequency of the $\mu$SR characteristic time window which is much faster than dynamic fluctuations of nuclear dipoles.\cite{Hayano} Accordingly, Eq.(\ref {eq_GLKT}) should be appropriate to analyze time spectra obtained from the $\mu$SR measurement on $\lambda$-(BETS)$_2$GaCl$_4$.

We applied Eq.(\ref {eq_GLKT}) to intermediate $\mu$SR time spectra measured in $\lambda$-(BETS)$_2$GaCl$_4$. Figure 8 is the best fit results done by using Eq.(\ref {eq_GLKT}). The time spectrum was measured at 1 K, 10 K, 20 K, and 50 K in which the system is in the paramagnetic state and the $\mu$SR time spectrum did not show the temperature dependence at all. The fitting results seem to be well successful with value of $a$ and $\Delta$ to be 0.10(1) $\mu$sec$^{-1}$ and 0.14(1)$\mu$sec$^{-1}$, respectively. This results indicates that the distribution of local fields at the muon site coming from surrounding nuclear dipoles deviates from Gaussian and becomes to be the intermediate shape. Since $\lambda$-(BETS)$_2$GaCl$_4$ has the anisotropic low-dimensional crystal structure, there are some spatial regions where the density of nuclear dipoles is largely different. In such a case, some muons which stop near the high- and low-density areas feel stronger and weaker local fields, respectively. This condition makes the field distribution wider and deforms the Gaussian shape. 

\begin{figure}[h]
\begin{center}
\includegraphics[width=15cm]{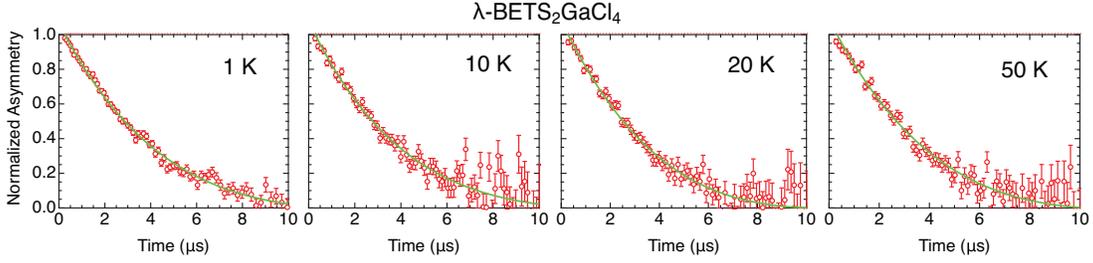}
\end{center}
\caption{Time spectra measured on $\lambda$-(BETS)$_2$GaCl$_4$ at 1 K, 10 K, 20 K and 50 K.\cite{Dita} Green solid lines indicates the best fit result by using  using Eq.(\ref{eq_GLKT}). }
\end{figure}

\subsection{Muon-spin depolarization by fluctuating dynamic local fields}

\ \ \ In addition to the static regime, the dynamic regime due to the appearance of fluctuating dynamic local fields also causes changes in the $\mu$SR time spectrum deforming its shape from Gaussian to the intermediate one as the function of the temperature. An example showing this case was obtained on the La-based high-$T_{\rm c}$ oxide, La$_{2-x}$Sr$_x$CuO$_4$ with $x$ of 0.024. In this Sr-doping regime, the system was underdoped of carriers and showed the magnetic transition around 10 K. Besides, the $\mu$SR time spectrum was found to start to divert from Gaussian below 100 K, forming the intermediate shape.\cite{IW0,IW6} Our previous study on this system used Eq.(\ref{eq_LG}) in order to discuss changes in the time spectrum on the basis of the appearance of effects of fluctuating dynamic local fields coming from surround electronic spins. Although the fitting of time spectra seemed to be good, the possibility of the trading-off effect between $\lambda$ and $\Delta$ could not be removed from the results and discussions. The similar behavior of the $\mu$SR time spectrum in the paramagnetic state was also reported in other high-$T_{\rm c}$ oxides,\cite{Panagopo,Son,SonPRB} so that the origin of this change in the $\mu$SR time spectrum in the paramagnetic state has been argued to be intrinsic to understand the mechanism of the high-$T_{\rm c}$ superocnductivity.\cite{Varma,Fauque,Xia} However, neither static nor dynamic properties of local magnetic fields which causes tiny changes in the $\mu$SR time spectrum has been clear due to the lack of the appropriate intermediate analysis function which can describe the time spectrum between Gaussian and Lorentzian ones. Following this situation, we applied Eq.(\ref{Dynamic_GLKT_Final}) to $\mu$SR time spectra measured in La$_{2-x}$Sr$_x$CuO$_4$ for $x$=0.024 and tried to reveal the dynamic and static properties of local fields at the muon site. In this case, we can recognize the fluctuating internal field at the muon site as the relative motion against the muon within the scheme of Eq.(\ref{Dynamic_GLKT_Final}). 

\begin{figure}[h]
\begin{center}
\includegraphics[width=15cm]{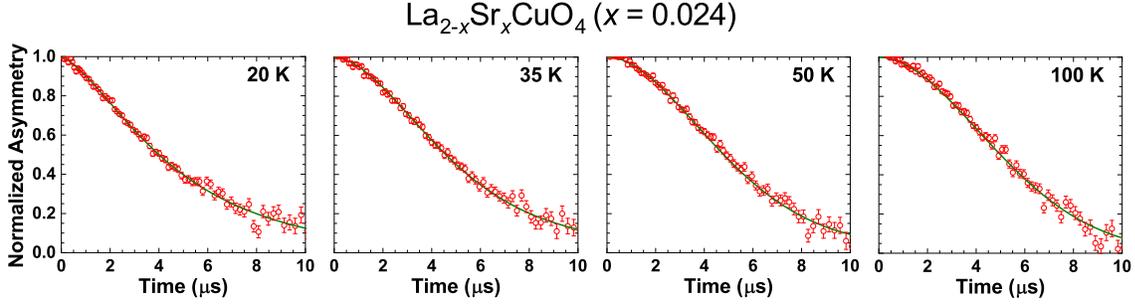}
\end{center}
\caption{Analysis results of time spectra measured in La$_{2-x}$Sr$_x$CuO$_4$ for $x$=0.024 at various temperatures. Solid lines in the figure indicate the best-fit results by using Eq.(\ref{Dynamic_GLKT_Final}).}
\end{figure}

\begin{figure}[h]
\begin{center}
\includegraphics[width=15cm]{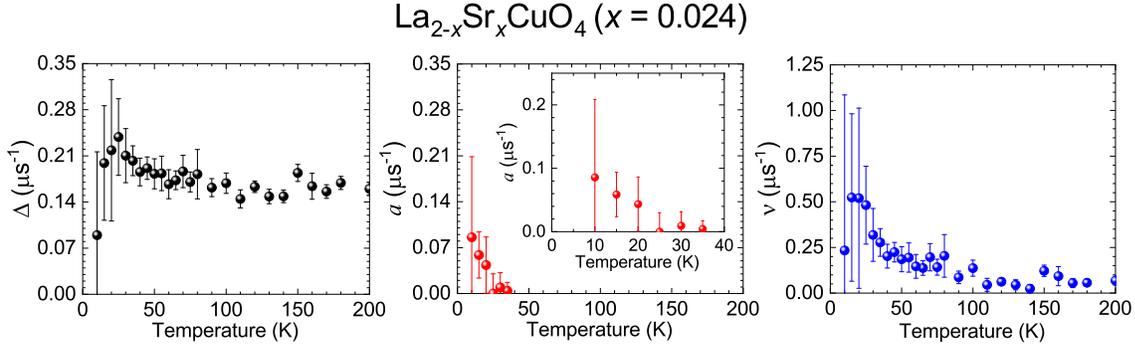}
\end{center}
\caption{Temperature dependences of $\Delta$ (left), $a$ (middle) and $\nu$(right) obtained by the application of Eq.(\ref{Dynamic_GLKT_Final}), respectively. Changes in $\Delta$ was different from that obtained by applying the phenomenological function of Eq.(\ref{eq_LG}).\cite{IW6}}
\end{figure}

Figure 9 shows the fitting results of some of $\mu$SR time spectra observed at about 20 K,  35 K, 50 K and 100 K in La$_{2-x}$Sr$_x$CuO$_4$ for $x$=0.024. Below 20 K, the temperature was too close to the magnetic transition temperature and the time spectrum becomes nearly the simple Exponential type reflecting that the fluctuating internal field from surrounding electrons became mandatory. As can be seen, the fitting results were successful proving that Eq.(\ref{Dynamic_GLKT_Final}) well worked to describe the intermediate state of local fields including fluctuating dynamic components. 

Figure 10 displays temperature dependences of $\Delta$ (left), $a$ (middle) and $\nu$(right), respectively. The present analysis by using Eq.(\ref{Dynamic_GLKT_Final}) demonstrates independent properties of each parameter. One new finding was that both $\Delta$ and $\nu$ increased below around 100 K at where the $\mu$SR time spectrum started to deviate from the Gaussian shape while $a$ still keeps to be almost nothing down to around 20 K. Especially, the temperature dependence of $\Delta$ was different from that obtained in our previous results.\cite{IW6} Therefore, we can finalize that changes below about 100 K in the time spectrum observed in La$_{2-x}$Sr$_x$CuO$_4$ for $x$=0.024 is not due to the trading-off effect between $\Delta$ and $\lambda$ but due to increase of both the width of the static Gaussian distribution and fluctuating internal fields at the muon site. 

\section{Summary}
\ \ \ We derived $\mu$SR relaxation functions under crossover magnetic fields between Gaussian and Lorentzian. We gave in this report a firm basis for matching the relaxation function parameters to the field distribution. Forms of those relaxation functions were found to be a kind of extension of the Kubo-Toyabe relaxation function. We succeed to describe their relaxation function of the muon-spin polarization which was in the intermediate state between Gaussian and Lorentzian in the zero-field and in-field cases.  

As demonstrations of our developed analysis equations, we applied them to real $\mu$SR data obtained in the organic molecular superconductor, $\lambda$-(BETS)$_2$GaCl$_4$, and the La-based high-$T_{\rm c}$ superconduting cuprates, La$_{2-x}$Sr$_x$CuO$_4$ for $x$=0.024 in which the intermediate $\mu$SR time spectrum was observed in the paramagnetic state. We have succeeded to reproduce time spectra by using our developed functions. This achievement can correct our previous data obtained from the applications of the phenomenological function of exp$(-\lambda t) \times P_{GKT}$ which would contain the trading-off effect between two parameters, $\lambda$ and $\Delta$. 

The current results and analysis equations described in this report can help to analyze the $\mu$SR data, and to discuss the physics outlook of the crossover and magnetic transition phenomenon in a clearer manner. As an example, when the ZF-$\mu$SR time spectrum deviates from the Gaussian shape in the crossover region and one can analyze the deviation by using Eq.(\ref {eq_GLKT}), the analysis result indicates the appearance of additional spontaneous internal fields which are comparable to $\Delta$ and characterized by $a$. Further more, it is definite that those additional fields are static from the view of the characteristic $\mu$SR time window. This analysis method should be worthwhile to quantitatively investigate those spontaneous small internal fields which are expected to result from exotic electronic properties of strongly correlated systems, such as a pseudo gap of the high-$T_{\rm C}$ superconducting oxides \cite{IW6,Panagopo} and the time reversal-symmetry breaking of the superconducting pairing symmetry.\cite{Luke}

\section*{Acknowledgment}
The authors are grateful to Mr. Muhammad Hanif Che Lah for his cooperative help. This study is supported by the Junior Research Associate (JRA) program of RIKEN and the JSPS KAKENHI (No 20H04463).





\vspace{0.2cm}
\noindent

\let\doi\relax
\bibliographystyle{ptephy}
\bibliography{GLfield}


%
%
%
%
%
%

\end{document}